\documentclass[twocolumn,showpacs,preprintnumbers,amsmath]{revtex4}
\usepackage{dsfont,amsmath,amssymb,graphicx}
\begin{document}
\title{Multiple Invaded Consolidating Materials}
 
\author{A. D. Ara\'ujo$^{1,2}$, J. S. Andrade Jr.$^{1,2}$ and H. J. Herrmann$^{1,2}$}
\affiliation{$^1$Departamento de F\'{\i}sica, Universidade Federal do 
Cear\'a, 60451-970 Fortaleza, Cear\'a, Brazil.\\
$^2$ Institut f$\ddot{u}$r Computeranwendungen $1$
Universit$\ddot{a}$t Stuttgart, 70569 Stuttgart, Germany.}

\date{\today}

\begin{abstract}
We study a multiple invasion model to simulate corrosion or
intrusion processes. Estimated values for the fractal
dimension of the invaded region reveal that the critical
exponents vary as function of the generation number $G$,
i.e., with the number of times the invasion process takes
place. The averaged mass $M$ of the invaded region decreases
with a power-law as a function of $G$, $M\sim G^{\beta}$,
where the exponent $\beta\approx 0.6$. We also
find that the fractal dimension of the invaded cluster
changes from $d_{1}=1.887\pm0.002$ to
$d_{s}=1.217\pm0.005$. This result confirms that the
multiple invasion process follows a continuous transition
from one universality class (NTIP) to another (optimal
path). In addition, we report extensive numerical simulations that
indicate that the mass distribution of avalanches $P(S,L)$
has a power-law behavior and we find that the exponent
$\tau$ governing the power-law $P(S,L)\sim S^{-\tau}$
changes continuously as a function of the parameter $G$. We
propose a scaling law for the mass distribution of
avalanches for different number of generations $G$.
\end{abstract} 

\maketitle

%%%%%%%%%%%%%%%%%%%%%%%%%%%%%%%%%%%
\section{Introduction}
%%%%%%%%%%%%%%%%%%%%%%%%%%%%%%%%%%%
The veins of gems and ores are often the product of a
multiple intrusion of a reacting fluid into a porous soil in
which dissolution and subsequent re-crystallization
processes are the determining factor. Some examples like
porphyry copper deposits \cite{Einandi_94} or olivine
\cite{Wanamaker_90} have been studied in the literature and
it is known that the surviving network of ore deposits has a
fractal structure \cite{Turcottei_92,Manning_94} that can be
exploited for mineral exploration \cite{Panabi_04}. A
similar situation can be found in vulcanology when magma is
repeatedly injected through the same pathway, each time
melting up again the most recent formations to find its way
out \cite{Luijten_97}.

The evolution of the pore structure after several
invasion-frost-thaw events has been investigated numerically
\cite{Salmon_97}, and results indicate that the fractal dimension
of invasion clusters varies with the number of invasion
cycles. In this work, after invasion takes place, the
structure of the porous pathway is randomly healed. In a
similar approach \cite{Zara_99}, an optimized version of the
multiple invasion percolation model was studied. Some
topological aspects as the acceptance profile and
the coordination number were investigated and compared to those
of ordinary invasion percolation.

In the cases mentioned
\cite{Einandi_94,Wanamaker_90,Turcottei_92,Manning_94,Panabi_04}
above and also in other cases of repeated invasions of
corroding, dissolving or melting fluids into a strongly
heterogeneous substrate, slowly consolidating matrix fractal
patterns are created that reflect the history of the
material. It is the aim of this paper to develop a model of
multiple invasion in order to simulate how these patterns
form and how their fractal dimension changes. In fact, we
propose a complete theoretical framework based on scaling
laws
\cite{Luijten_97}.

The theory of avalanche dynamics has been studied in a
variety of contexts, for example in growth models,
interface depinning and invasion percolation
\cite{Paczuski_96}.  The formation of fractal structures, 
diffusion with anomalous Hurst exponents and L\'evy flights, can
all be related to the same underlying avalanche dynamics
\cite{Paczuski_96}. Normally, the presence of avalanches 
in the invasion process supposes unchanged porous media. In
this work we also investigate the mass-distribution of
avalanches and determine how the exponent that characterizes
this distribution changes for different cycles of the
invasion process. This paper is organized as follows. In
Sec. II, we present the model to simulate the multiple
invasion in consolidating medium. In Sec. III we show the
results for the invaded cluster mass. The results and
analysis of the numerical simulation for avalanche
distribution are shown in Sec. IV., while the conclusions
are presented in Sec. VI.

%%%%%%%%%%%%%%%%%%%%%%%%%%%%%%%%%%%%%
\section{Model}
%%%%%%%%%%%%%%%%%%%%%%%%%%%%%%%%%%%%%
In order to simulate the injection process we use the
standard non-trapping invasion percolation (NTIP)
\cite{Wilkinson_83}. In this model the invaded solid
is considered to be very heterogeneous and the invading
fluid can potentially enter anywhere along the
interface. Here the consolidating medium is represented
conveniently as a square network. The sites of the lattice
can be viewed as the smallest units of constant strength and
the randomness of the strength of the medium is incorporated
by assigning random numbers to sites. For simplicity, we
consider the case in which dissolutions control the fluid
invasion.

On our heterogeneous medium we start by applying the
standard invasion process of NTIP. For completeness the
algorithm is described as follows. Initially, let us assign a
random number, $p_{i}$ drawn from a uniform distribution
in the interval $[0,1]$, to each site $i$ of the lattice. We
choose one site in the center of the lattice and occupy
it. This site represents the injection point of the fluid
and is the seed of the invading cluster. We look among the
neighboring sites of this cluster (the growth sites) and
choose the one which carries the smallest random
number. This site is then invaded and added to the
cluster. Then we increase the list of sites that are
eligible to be invaded. At each step of the invasion
process, the perimeter of the nearest neighbors of the sites
that form the invading cluster is investigated and the site
that has the smallest $p_i$ is chosen. This procedure is
repeated until the edge of the lattice is reached. At this
point the simulation stops and the mass $M$ (i.e. , the
number of sites belonging to the invaded cluster) of the
cluster is computed. The number of sites of the invaded
cluster is very often considered as a {\it time} parameter.
\begin{figure}
\begin{center}
\includegraphics[width=3.0cm]{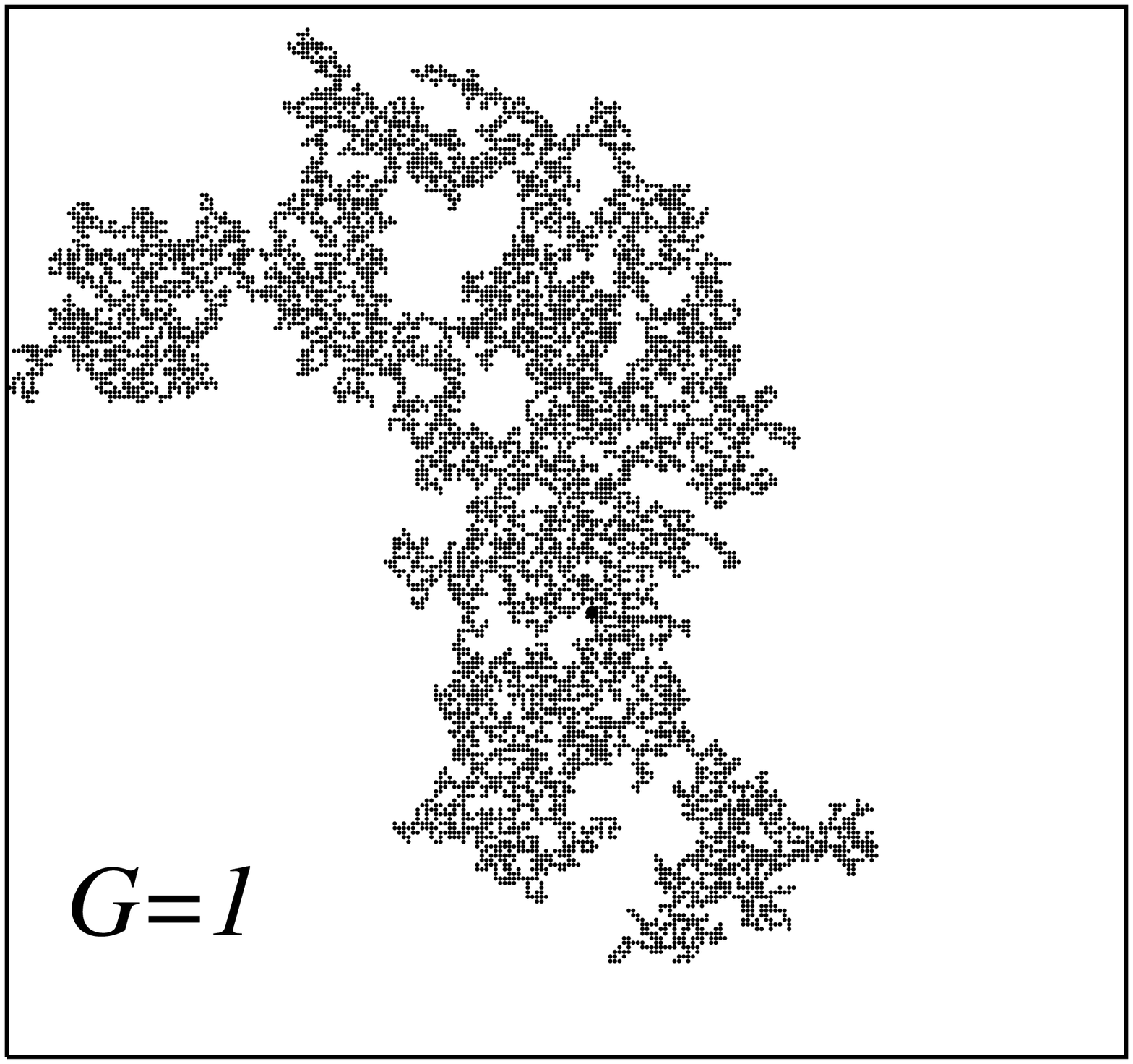}
\includegraphics[width=3.0cm]{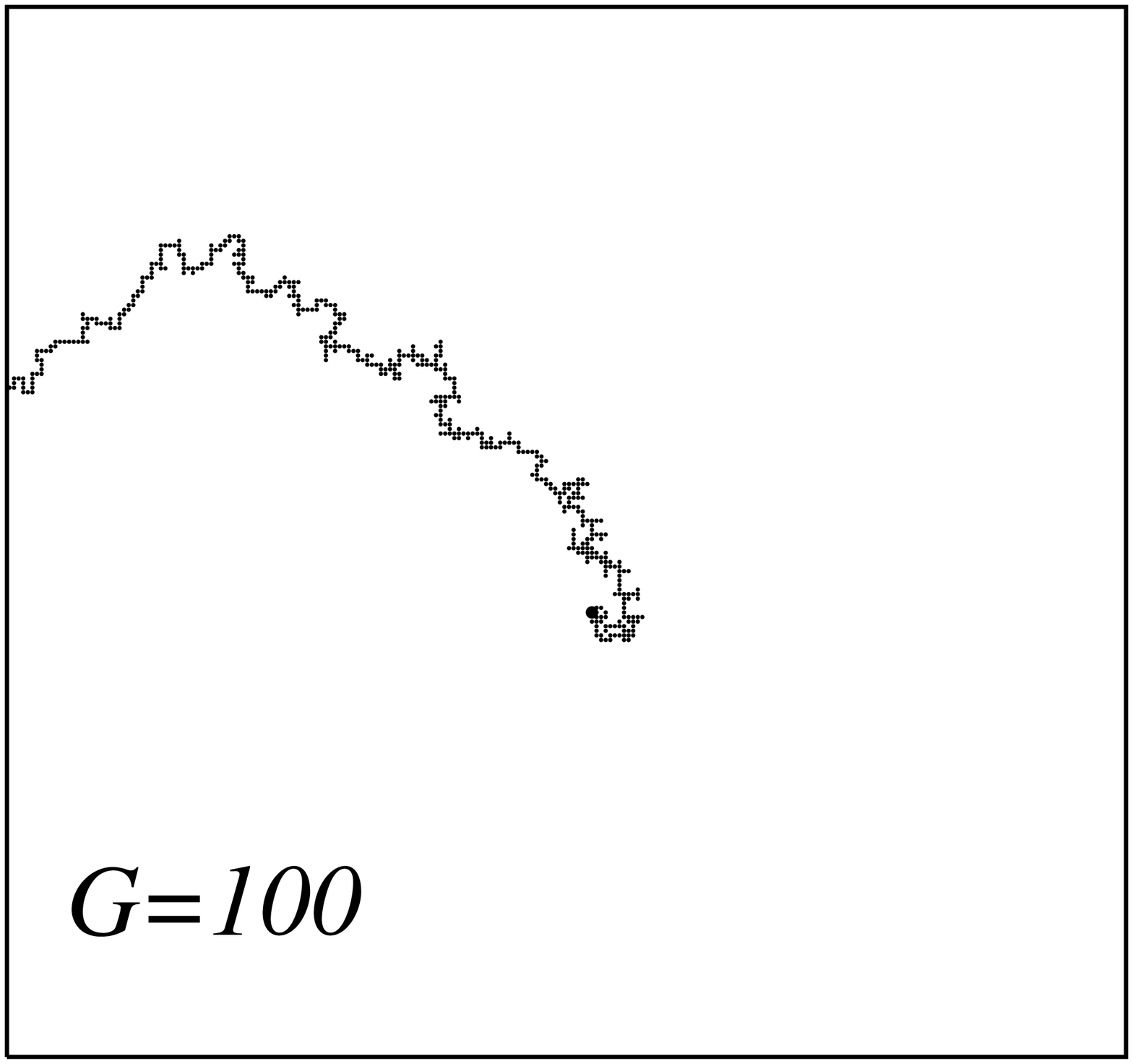}
%\vskip 0.5cm
\caption{Typical cluster for different 
generations on a $256\times256$ lattice, for $p^{*}=1$. The
injection point is localized in the center of the
lattice. In figures we show the results for $G=1$ and $100$.}
\label{fig_geometry}
\end{center}
\end{figure}

Now we present the new feature introduced to the standard
invasion percolation. After we finish the above described
simulation in agreement with customary NTIP, the simulation
is performed again starting every time at the same injection
point. New random numbers chosen from a uniform distribution
in the interval $[0,1]$ are assigned to all sites
belonging to the previously invaded cluster before a new
invasion process starts. To all other sites, i.e., namely,
those that are outside the cluster, we assign a random
number homogeneously distributed in the interval
$[p^{*},1]$ where $p^{*}$ is a number close to
unity. Compared to the support used in the first generation
where all sites can be invaded, the second generation
appears substantially reduced, because it mostly corresponds
to the cluster invaded in the first generation. In this way
we generate again an invasion cluster for which $p^{*}=1$
is a subset of the previous one and so necessarily
smaller. This procedure is repeated $G$ times, where $G$
is the number of generations. Standard invasion percolation
coincides with the case $G=1$. At each new generation
$G$, the sites of the previous invasion are re-invaded.

Let us first consider the case $p^{*}=1$. In this
situation, the invaded cluster is after each time a subset
of the previous one so that after a finite number of
iterations the cluster does not decrease any longer. The
number of generations needed to reach a cluster whose mass
remains unchanged depends on the size of the original
lattice, because the number of possible available sites is
proportional to the system size. Therefore, the saturation
number is different for each lattice size. In
order to illustrate these changes in the structure after
each process of invasion, we show in Fig.~\ref{fig_geometry}
typical clusters generated for a lattice of size $L=256$
for four different generations $G$.
\begin{figure}
\begin{center}
\includegraphics[width=8.0cm]{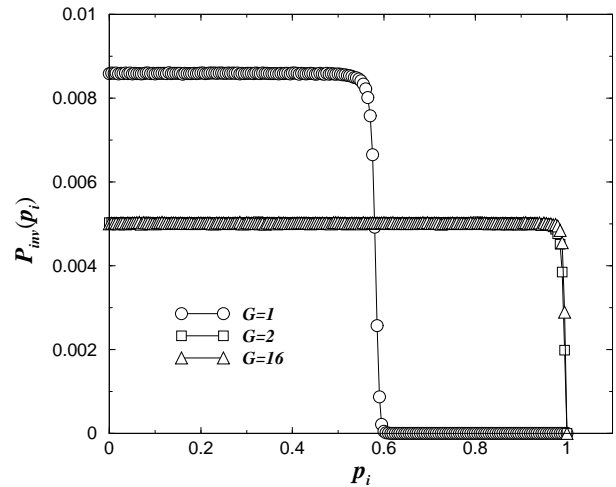}
\vskip 0.5cm
\caption{The probability distribution $P_{inv}(p)$ of invaded sites for
different generations $G=1$ (circles), $2$(squares) and
$16$ (triangles), $L=512$ and $p^{*}=1$.}
\label{fig_hist_p}
\end{center}
\end{figure}
Another important quantity is the probability distribution
of $p_{i}$ of the invaded sites. In Fig.~\ref{fig_hist_p} we
present the normalized distribution $P_{inv}(p)$ for
different generations $G$ obtained from $1000$ realizations
of size $L=512$. After the completion of the first invasion
process, the distribution expectedly displays a transition
at $p\approx p_{c}$, where $p_{c}$ is the critical site
percolation point, $p_{c}=0.59275$ for a square lattice
\cite{Stauffer_94}. The same behavior has been observed by
numerical simulation in Refs.~\cite{Salmon_97,Zara_99}. For
$G=2$ the distribution $P_{inv}(p)$ becomes flat and the
profile does not change any more as function of $G$. This
happens because when $G>1$ sites with larger $p_{i}$ are
also invaded.

%%%%%%%%%%%%%%%%%%%%%%%%%%%%%%%%%%%%%%%%%
\section{Cluster Mass}
\begin{figure}
\begin{center} 
\includegraphics[width=8.0cm]{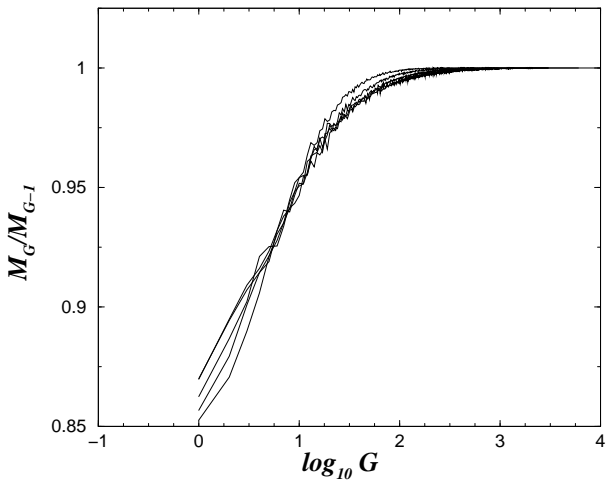}
\vskip 0.5cm
\caption{The evolution of the rate $M_{G}/M_{G-1}$ as function 
of the logarithm of the generation number $G$ for $p^{*}=1$. Here $M_G$
is the mass at generation $G$ for different sizes
$L=64,128,256$ and $512$.}
\label{fig_evolu}
\end{center}
\end{figure}

\begin{figure}
\begin{center} 
\includegraphics[width=8.0cm]{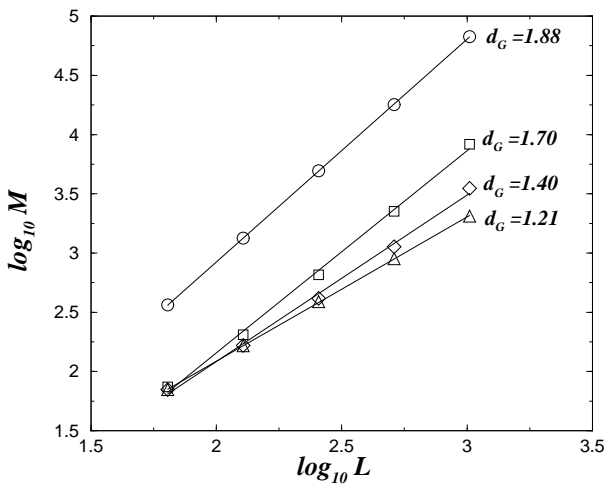}
\vskip 0.5cm
\caption{Log-log plot of the mass $M$ of the invaded cluster 
versus the system size for different generation numbers
$G=1$ (circles), $100$ (squares), $500$ (diamonds) and
$3000$ (triangles), and $p^{*}=1$. The straight lines are
best fits to the data and their slopes are the fractal
dimensions of the invaded clusters. }
\label{fig_dmf}
\end{center}
\end{figure}

\begin{figure}
\begin{center} 
\includegraphics[width=8.0cm]{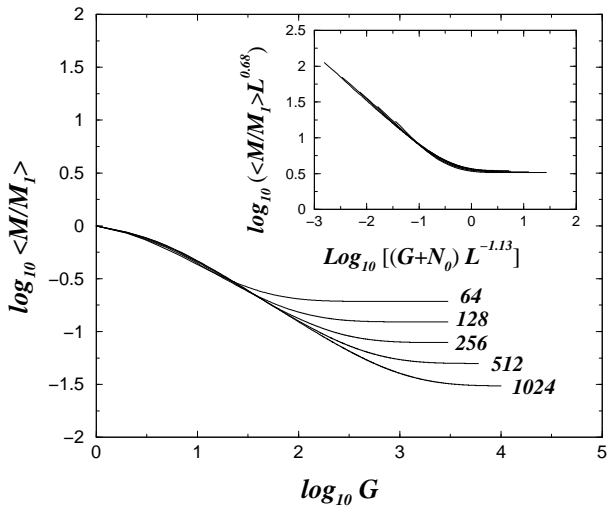}
\vskip 0.5cm
\caption{Log-log plot of the average mass of the invaded cluster $M$ 
normalized by the mass $M_1$ of the first invaded cluster,
against the number of generations $G$ for different sizes
$L=64,128,256,512$ and $1024$, and $p^{*}=1$. The inset
shows the collapse following the scaling relation of
Eq.~(\ref{eq_scaling}).}
\label{fig_ng_norm}
\end{center}
\end{figure}

%%%%%%%%%%%%%%%%%%%%%%%%%%%%%%%%%%%%%%%%%
In our simulations we used the NTIP algorithm for square
lattices of sizes $L=64,128,256,512$ and $1024$. For each
value of $G$, we perform simulations for $10000$
realizations and compute the mass $M_{G}$ of the invaded
cluster. In Fig.~\ref{fig_evolu}, we show the ratio
$M_{G}/M_{G-1}$ as a function of the generation number
$G$. For each size $L$, $G_{s}$ is defined as the number of
generations at which the mass of the invaded cluster reaches
a constant value, i.e., for which
$M_{G_{s}}/M_{G_{s}-1}=1$.
The results of our simulations shown in Fig.~\ref{fig_dmf}
for four values of the generation number $G$, indicate that
the mass $M$ has a power-law dependence on the size $L$,
$M\sim L^{d_{G}}$, where $d_G$ is the fractal dimension of
the invaded cluster. The case $G=1$ corresponds to the
standard invasion percolation model. The value obtained from
our simulations, $d_{1}=1.887\pm0.002$, is in good agreement
with the current estimate $d_{1}=1.8959$ for NTIP 
\cite{Feder_86,Stauffer_94,Schwarzer_99,Knackstedt_02}. 
The results shown in Fig.~\ref{fig_dmf} indicate that by
increasing the generation number the fractal dimension
decreases continuously until it reaches a saturation value
of $d_{s}=1.217\pm 0.005$ at $G_{s}$. This value agrees
with the fractal dimension of the optimal path in the strong
disorder limit $d_{opt}=1.22\pm 0.01$
\cite{Cieplak_94}.
As shown in Fig.~\ref{fig_ng_norm} for large system sizes we
find that the average mass of the invaded cluster
asymptotically follows a power-law behavior
\begin{equation}
M\sim G^{\beta}.
\label{eq_power}
\end{equation}
To better analyze the
data, we normalize the mass by the constant $M_{1}$, which
is the average mass of the invaded cluster at
$G=1$. Similar to some problems that involve growth
surfaces \cite{Barabasi_95}, this process has two
characteristic regimes: (i) power law evolution and (ii)
saturation when $G\rightarrow\infty$. To describe this
behavior we propose the scaling relation \cite{Luijten_97}
\begin{equation}
\frac {M(G,L)}{M_{1}}=L^{\alpha} f\left 
( \frac{G-N_{0}}{L^{z}} \right ).
\label{eq_scaling}
\end{equation}
where $N_{0}$ is an off-set value for the generation
number and $\alpha$ and $z$ are scaling exponents. We assume
that the scaling function $f(x)$ has the form $f(x)\sim
x^{\beta}$ in the limit $x\ll1$ and $f(x)=const$ when
$x\gg1$. Furthermore, a direct relation among exponents
$\alpha$, $\beta$ and $z$ can be obtained. We find
$M/M_{1}\sim G^{\beta}$ for $L>>1$ and, since
$M/M_{1}\sim L^{\alpha}$ in the saturation regime
$(G>>1)$, we obtain that $\alpha=d_{s}-d_{1}$.

In the  crossover region, when the fractal dimension goes
from $d_{1}$ $(G=1)$ to $d_{s}$ $(G=G_{s})$ we have 
\begin{equation}
(G-N_{0})\sim L^{z} .
\label{eq_cross}
\end{equation}
From these relations, we obtain that
\begin{equation}
z=\frac{\alpha}{\beta},
\label{eq_relexp}
\end{equation}
and from the fact that the fractal dimension has reached the
saturation value $d_{s}\approx1.22$, it gives $\alpha=-0.68$.
The inset of Fig.~\ref{fig_ng_norm} shows the data collapse
obtained by rescaling $M/M_{1}$ and $G$ according to the
scaling form Eq.~(\ref{eq_scaling}). In this case the best
fit to the data gives $\beta\approx 0.6$. Substituting into
Eq.~\ref{eq_relexp} we find $z=1.13$.

%

%%%%%%%%%%%%%%%%%%%%%%%%%%%%%%%%%%%%%%%%%
\section{Avalanche Distribution}
%%%%%%%%%%%%%%%%%%%%%%%%%%%%%%%%%%%%%%%%%

It has been known since a long time that avalanches occur in
invasion percolation and that these avalanches obey
scaling relations related to percolation theory
\cite{Roux_89}. An avalanche occurs when a site $j$ is
invaded at a value $p_{j}$ and then a series of sites
$i$ connected to this original site are sequentially
invaded with $p_{i}<p_{j}$. It is also known that the
system reaches a self-organized critical state characterized
by avalanches of all sizes distributed according to a power
law. In the case of NTIP, the exponent corresponding to the
power law behavior for the distribution $P(S)$ of
avalanche sizes $S$ is $\tau=1.527$ \cite{Roux_89}.

In our simulation we found that the exponent corresponding
to the case $G=1$ is $\tau=1.46\pm0.03$. The expected value
\cite{Roux_89} is outside of our error bars, which we attribute
to the fact that we have not reached the asymptotic limit
because our systems are too small.

\begin{figure}
\begin{center} 
\includegraphics[width=8.0cm]{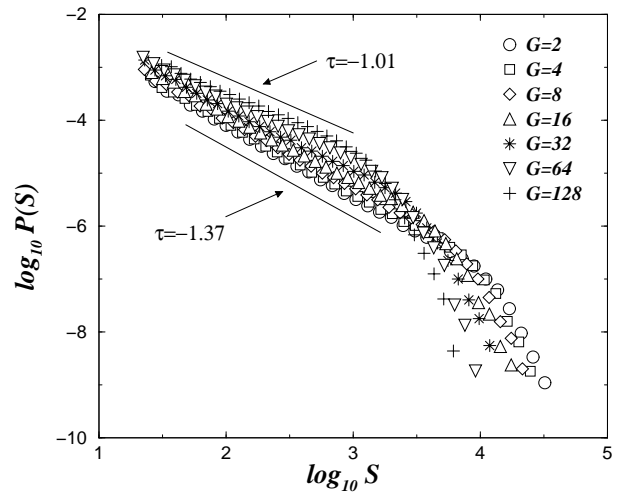}
\vskip 0.5cm
\caption{The mass distribution of avalanches for different
generation numbers $G=2,4,8,16,32,64$ and $128$ for
$L=512$ and $p^{*}=1$. The slopes of the straight lines
follow power-laws with exponent $\tau$. The solid lines
indicate the two limit cases $G=1$(lower) and $G=128$
(upper).}
\label{fig_ng_aval_512}
\end{center}
\end{figure}

We performed simulations for different generations $G$ on
lattices of sizes $L=64,128,256,512$ and $1024$, and
calculated the size distribution of avalanches. In
Fig.~\ref{fig_ng_aval_512} we show $P(S)$ for size
$L=512$ and $G=2,4,8,16,32,64$ and $128$. It is clear
from this figure that $P(S)$ displays power-law behavior
with exponent dependent of the number of generations
$G$. The solid lines indicate the slopes in the two limit
cases $G=1$ (lower) and $G=128$ (upper).

\begin{figure}
\begin{center} 
\includegraphics[width=8.0cm]{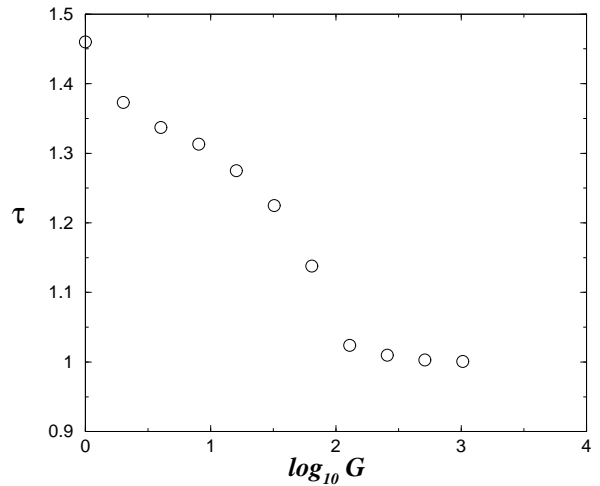}
\vskip 0.5cm
\caption{Log-linear plot of the avalanche exponent $\tau$ as function of the
generation number $G$, for $L=512$ and $p^{*}=1$.}
\label{fig_expava_ng}
\end{center}
\end{figure}

In Fig.~\ref{fig_expava_ng}, we show how the exponent of the
power-law $\tau$ changes as function of the number of
generations $G$. For large values of $G$ the exponent
converges to $\tau=1$. This value is the same found for the
distribution of avalanches $P(S)$ in the one-dimensional
case
\cite{Alencar_03}. This is consistent with
Fig.~\ref{fig_geometry} for $G=100$ where the avalanche process
is limited to a thin path that is essentially an
one-dimensional topology.

\begin{figure}
\begin{center} 
\includegraphics[width=8.0cm]{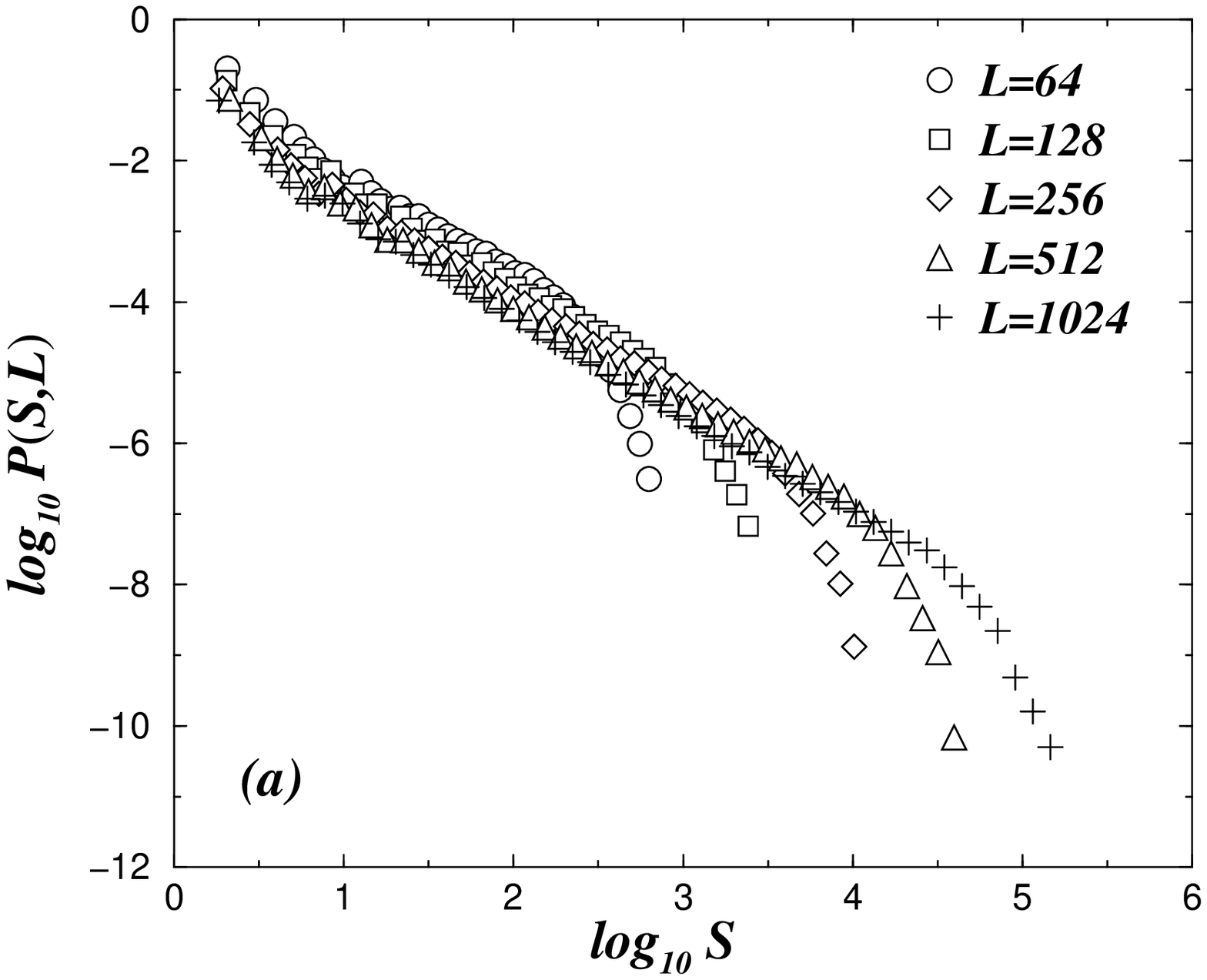}
\includegraphics[width=8.0cm]{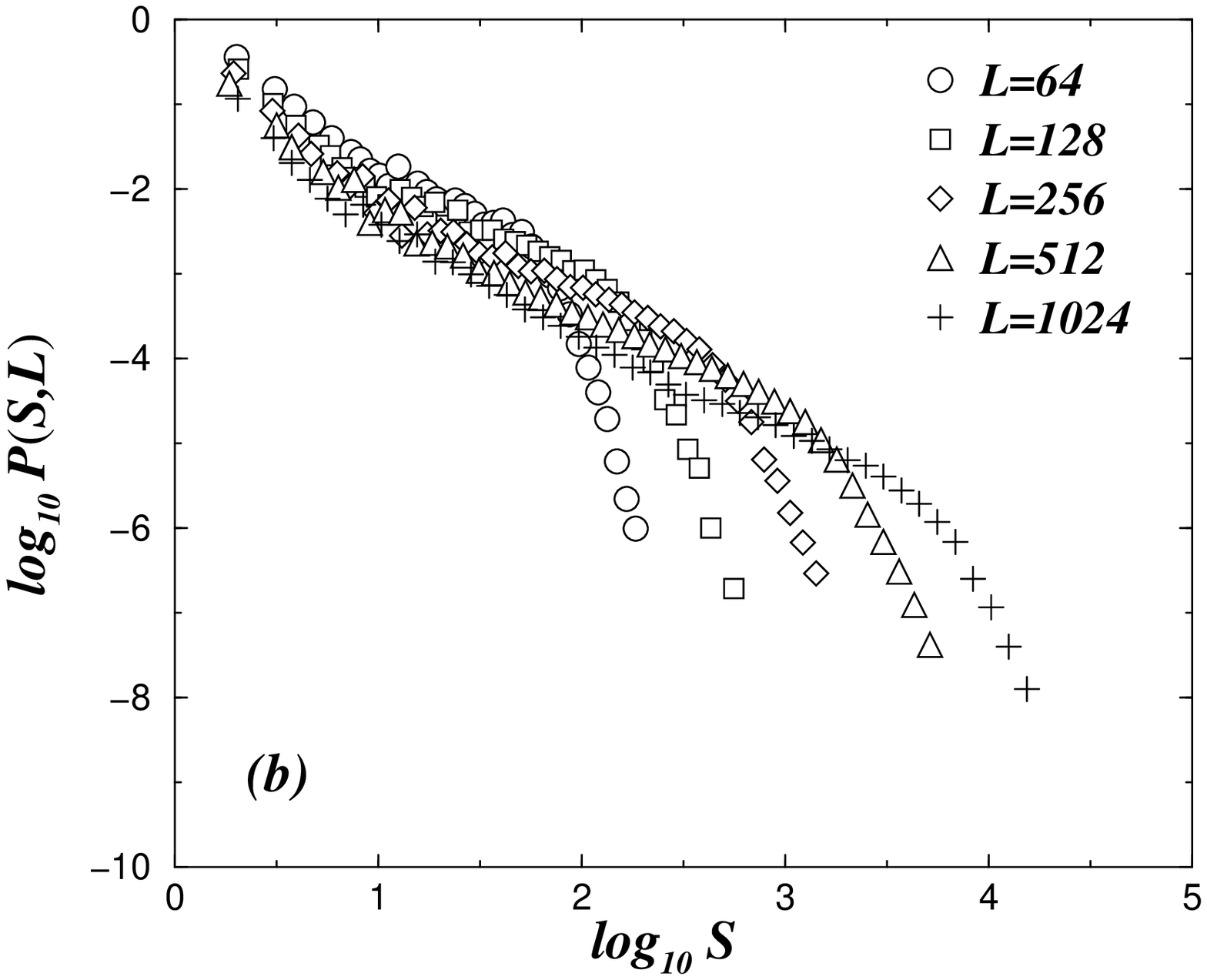}
\vskip 0.5cm
\caption{Log-log plot of the probability distribution of 
avalanches $P(S,L)$ for various sizes
$L=64,128,256,512$ and $1024$, $p^{*}=1$.(a) $G=2$ and (b)
$G=128$.}
\label{fig_P_SL}
\end{center}
\end{figure}

In Figs.~\ref{fig_P_SL}(a) and \ref{fig_P_SL}(b) we show the
log-log plot of the distribution of avalanche sizes. It is
clear from these figures that $P(S)$ displays a scaling
region for intermediate avalanche sizes. In addition the
scaling region is followed by a sudden cut-off that decays
faster than exponential due to a finite size
effect. The range of the power-law region is proportional to
the lattice size. As a consequence the biggest avalanches
occur in the largest lattice. The position of the cutoff
depends on $G$ for fixed $L$. We propose a scaling form
for the mass distribution $P(L,S)$, which accounts for
finite size effects and power-law behavior
\cite{Andrade00}
\begin{equation}
P(S,L)\propto S^{-\tau} f \left (\frac{S}{L^{\gamma}} \right ) ,
\label{scal_avalan}
\end{equation}
where the function $f(x)$ has a Gaussian form 
\begin{equation}
f(x)=exp[-x^{2}].
\end{equation}
\begin{figure}
\begin{center} 
\includegraphics[width=8.0cm]{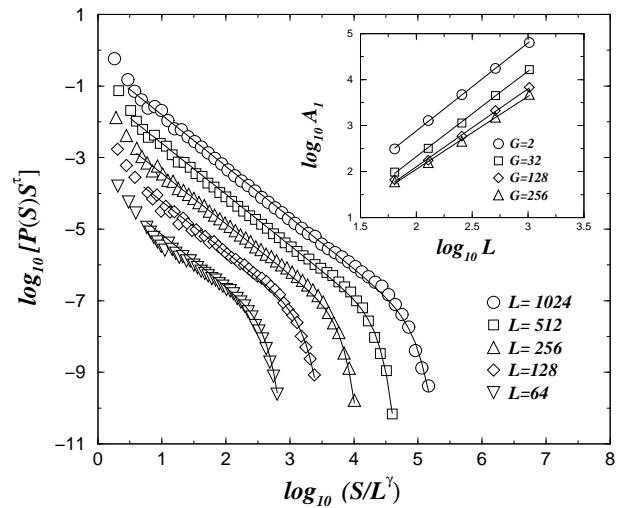}
\vskip 0.5cm
\caption{Log-log plot of the distribution $P(S)$ 
for $p^{*}=1$ and generation number $G=2$ for $L=1024$ (circles), $512$
(squares), $256$ (up triangles), $128$ (diamonds), and $64$
(down triangles). The solid lines correspond to the scaling
function $y=A_{0} S^{-\tau} exp[-(S/A_{1})^{2}]$ with the
parameter $\tau=1.37$. The inset shows the log-log plot of
the crossover amplitude $A_{1}$ versus the system size
$L$ for $G=2$ (circles), $32$ (squares), $128$ (diamonds), and
$256$ (triangles). The lines are the
least-square fits to the data and the slope is $\gamma$.}
\label{fig_fit}
\end{center}
\end{figure}
In practice, the appropriate parameters of the scaling
function Eq.~(\ref{scal_avalan}) have been determined here
through a nonlinear fitting procedure of the function
\begin{equation}
P(S,L)=A_{0} S^{-\tau} exp[-(S/A_{1})^{2}]
\label{equa_fit}
\end{equation}
to the avalanche data. We observe that both the prefactor
$A_{0}$ and the crossover amplitude $A_{1}$ depend on the
system size.

The solid line in Fig.~\ref{fig_fit} corresponds to the best
fit using Eq.~(\ref{equa_fit}) for $G=2$ and many
different sizes $L$ with $\tau=1.37$. The inset of
Fig.~\ref{fig_fit} shows the power-law dependence of
crossover amplitude on the system size, $A_{1}\propto
L^{\gamma}$. The straight lines are the least-square fits to
the data, with the slopes corresponding to the exponent
$\gamma$ in Eq.~(\ref{scal_avalan}) for different
generation numbers.

\begin{figure}
\begin{center} 
\includegraphics[width=8.0cm]{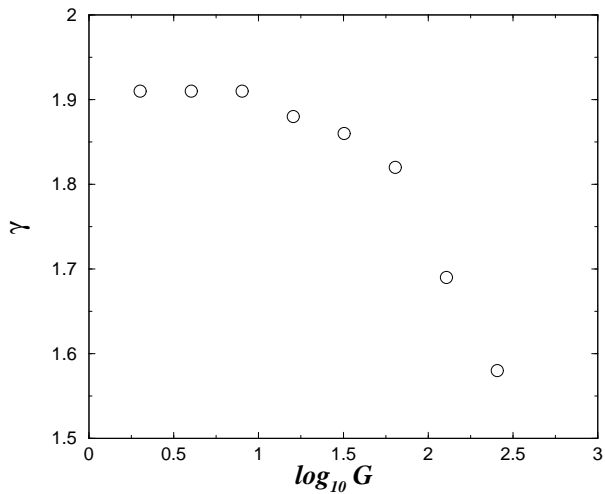}
\vskip 0.5cm
\caption{Log-linear plot of the exponent $\gamma$ versus
the generation number $G$. $p^{*}=1$.}
\label{fig_expgama}
\end{center}
\end{figure}

In Fig.~\ref{fig_expgama} we plot the exponent $\gamma$
versus $G$, and see that the exponent has a monotonic
behavior as function of the generation number.

\begin{figure} 
\begin{center} 
\includegraphics[width=8.0cm]{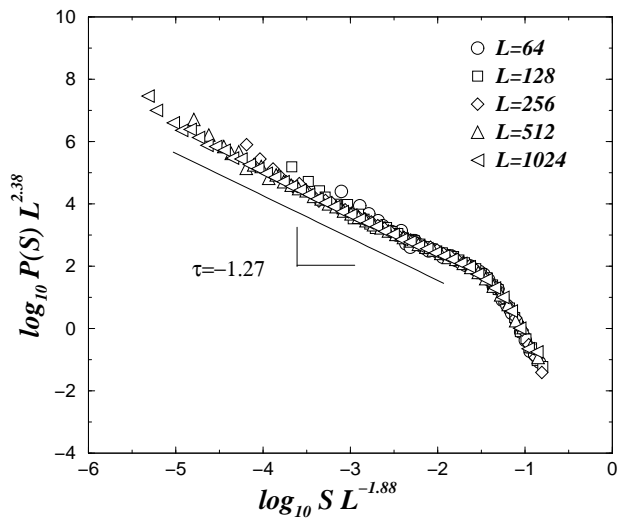}
\vskip 0.5cm
\caption{Log-log plot of the rescaled distribution 
of avalanches sizes $P(S/L^{\gamma})$ for generation
number $G=16$ and different lattice sizes
$L=64,128,256,512$ and $1204$ and $p^{*}=1$.}
\label{fig_scal1_ng}
\end{center}
\end{figure}

In Fig.~\ref{fig_scal1_ng} we show the rescaled function
$P(S/L^{\gamma})$ for $G=16$. The data collapse obtained
confirms the validity of the scaling form of
Eq.~(\ref{equa_fit}). This confirms that the system is
self-organized critical and the rescaled distribution shows
the asymptotic scaling behavior of
Eq.~(\ref{equa_fit}).

%%%%%%%%%%%%%%%%%%%%%%%%%%%%%%%%%%%%%%%%%
\section{Results for  $p^{*}\neq1$}
%%%%%%%%%%%%%%%%%%%%%%%%%%%%%%%%%%%%%%%%%

\begin{figure}
\begin{center} 
\includegraphics[width=8.0cm]{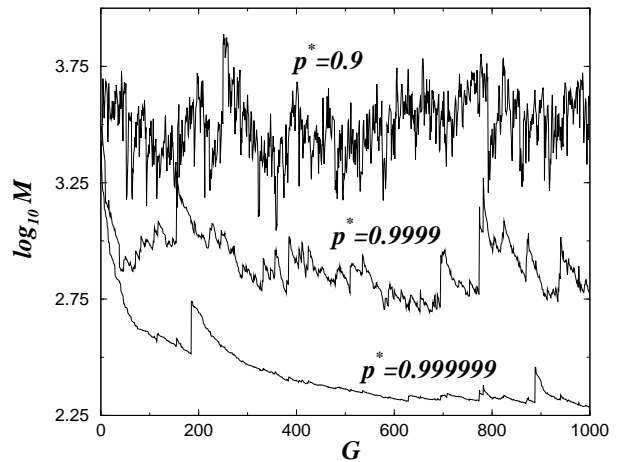}
\vskip 0.5cm
\caption{Linear-log plot of the invaded mass for a 
typical realization as function of the generation number
$G$, for $L=512$. From top to bottom,
$p^{*}=0.9,~0.9999,~0.999999$.}
\label{fig_mass_p*_ng}
\end{center}
\end{figure}

\begin{figure}
\begin{center}
\includegraphics[width=3.0cm]{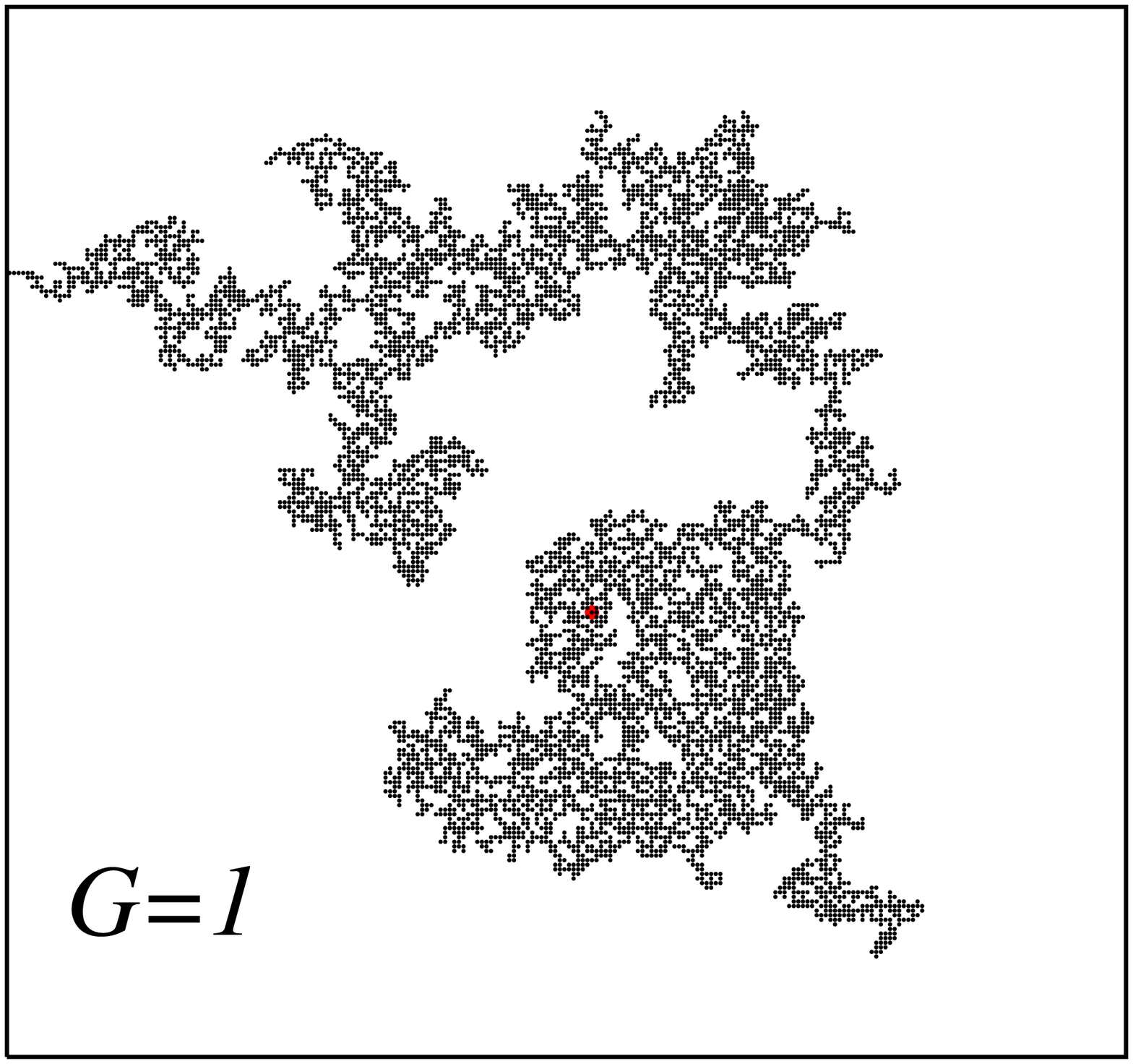}	
\includegraphics[width=3.0cm]{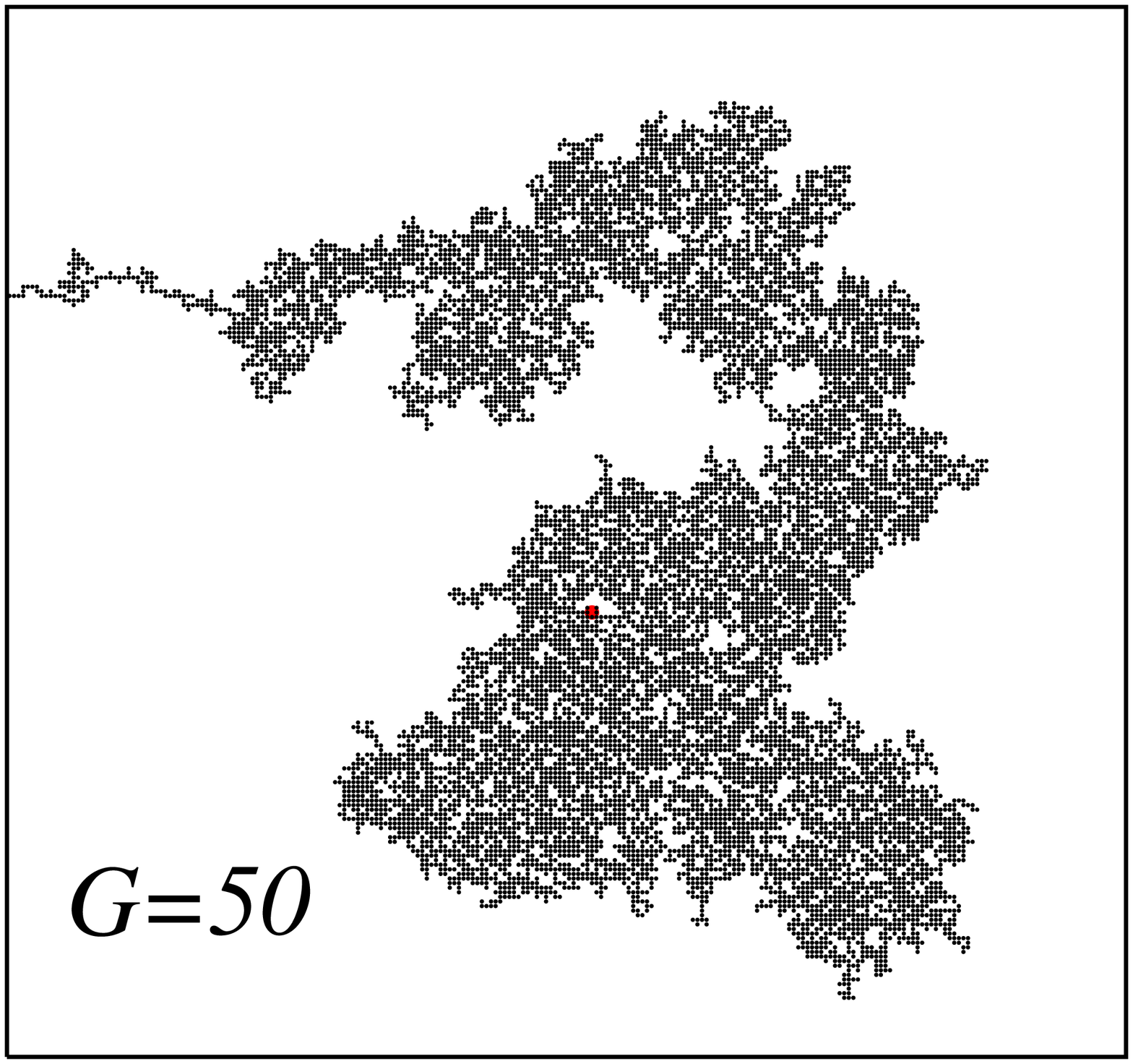}
%\vskip 0.5cm
\caption{Typical configurations of invaded clusters 
at different generations $G$ and $L=256$. The random
number, $p_i$ drawn from a uniform distribution of
probabilities in the interval $[p^*,1]$ for $p^*=0.9$.}
\label{clus_1}
\end{center}
\end{figure}

\begin{figure}
\begin{center}
\includegraphics[width=3.0cm]{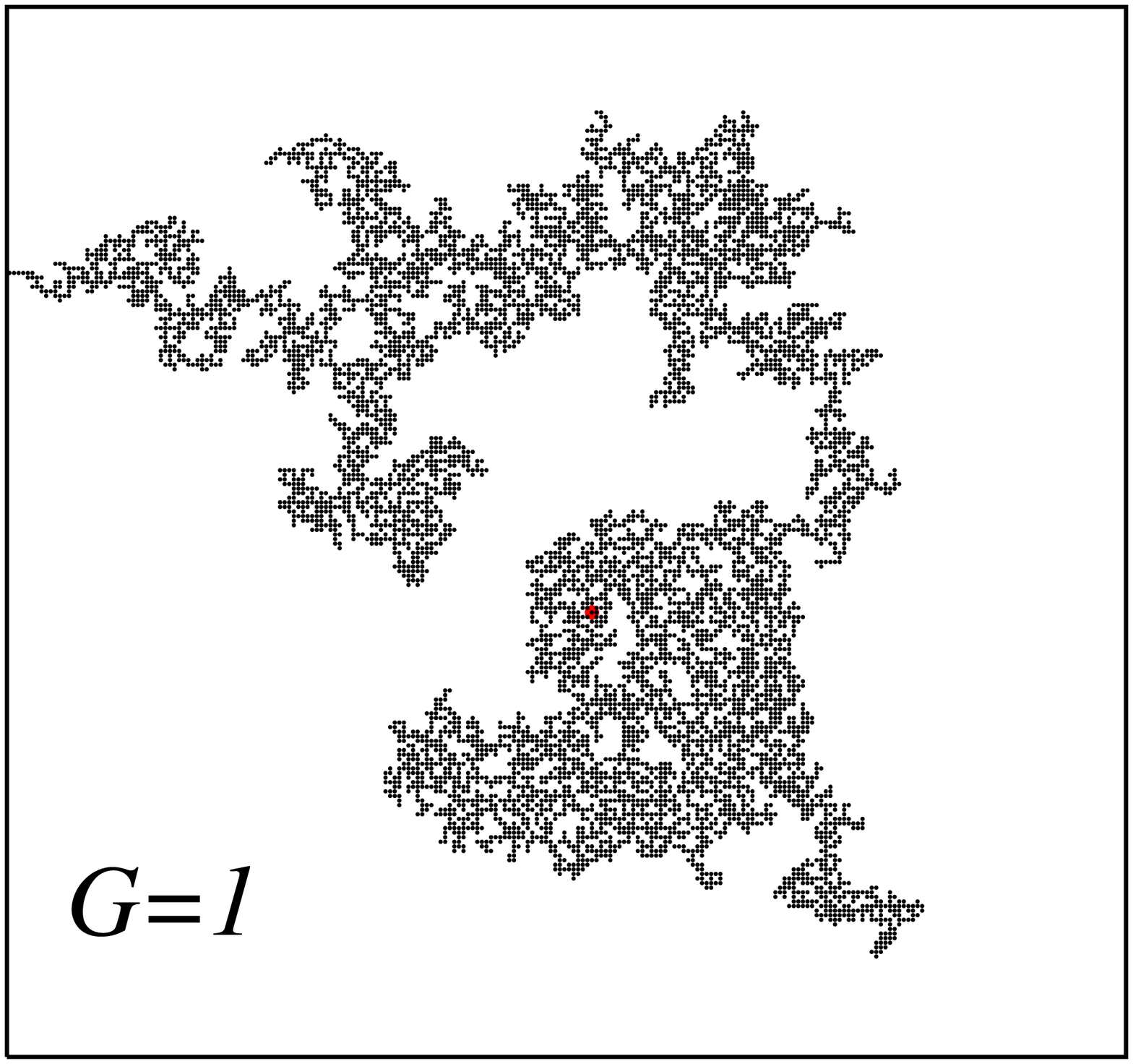}
\includegraphics[width=3.0cm]{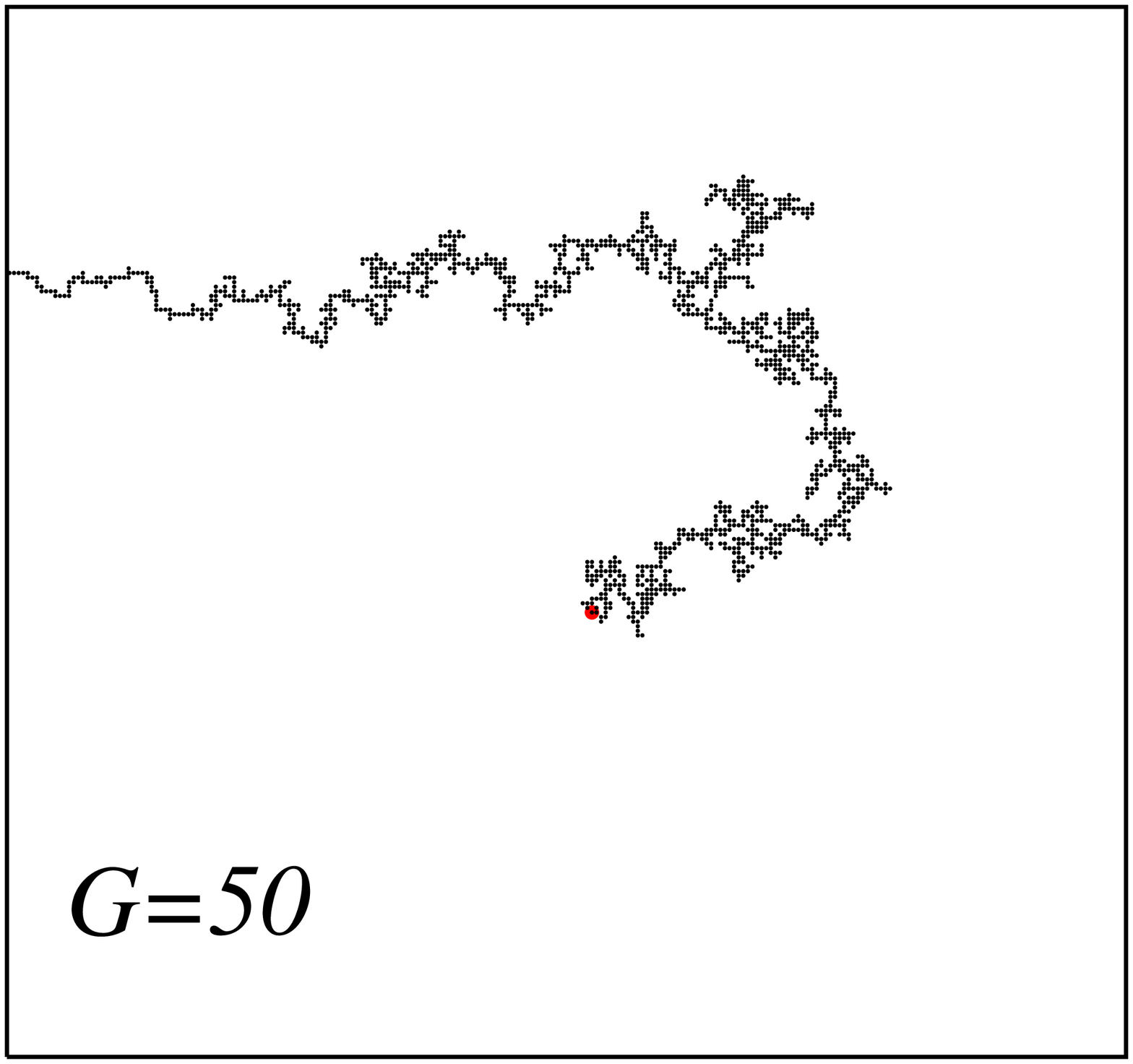}
%\vskip 0.5cm
\caption{Typical cluster configurations for invaded clusters 
at different generations $G$ and $L=256$. The random
number, $p_i$ drawn from a uniform distribution of
probabilities in the interval $[p^*,1]$ for $p^*=0.999999$.}
\label{clus_2}
\end{center}
\end{figure}

In the first part of this work we considered $p^{*}=1$. Now
we present simulations for different $p^*$ very close to
unity. In Fig.~\ref{fig_mass_p*_ng} we show how the mass of
the invaded cluster varies as function of the generation
number $G$ for a typical realization of the multiply
invasion process. In the case $p^{*}=0.9$ the value of the
mass shows strong fluctuations. If the probability to occupy
sites outside of the previously invaded cluster is raised,
the previous invaded region of the porous media is more
likely invaded.
To understand the qualitative behavior of the invaded
cluster as function of the generation $G$ we show in
Figs.~\ref{clus_1} and ~\ref{clus_2}, typical clusters for two
values $p^*=0.9$ and $p^*=0.999999$, for five different
generations $G=1,5,10,25$ and $50$. For $p^*=0.9$, the cluster
is more compact and sometimes changes the point where it
reaches the border. When $p^*=0.999999$, the cluster
becomes smaller at each generation.

\begin{figure}[ht]
\begin{center}
\includegraphics[width=8.0cm]{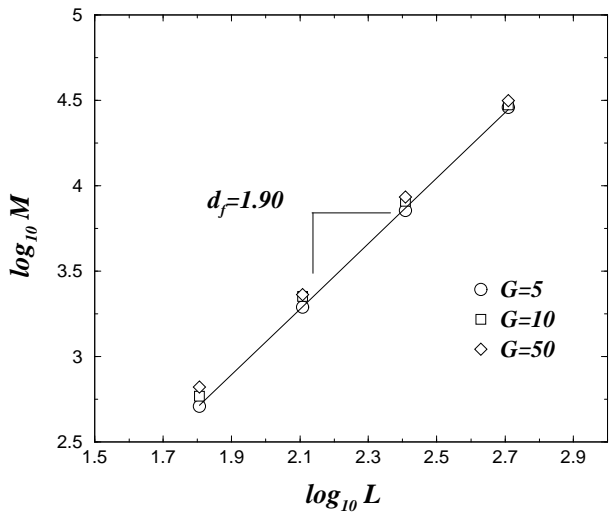}
\vskip 0.5cm
\caption{Log-log plot the averaged mass	$M$ as function of the system size 
$L$ for $p^{*}=0.9$ and $G=5$ (circles),
$10$ (squares) and $50$ (diamonds). The solid line with
slope $1.90\pm 0.01$ is the least-square fit to all data
sets.}
\label{dmf_p0.9}
\end{center}
\end{figure}

\begin{figure}[ht]
\begin{center}
\includegraphics[width=8.0cm]{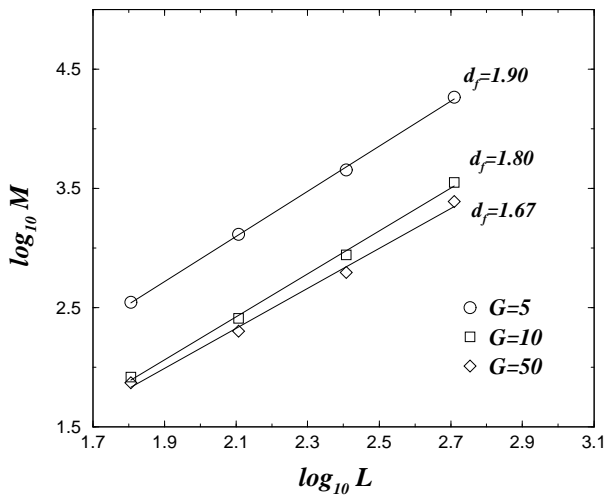}
\vskip 0.5cm
\caption{Log-log plot of the averaged mass $M$ against the 
system size $L$ for $p^{*}=0.999999$ and $G=5$ (circles),
$10$ (squares) and $50$ (diamonds). The straight
lines are least-square fits to the data, with the numbers
corresponding to the fractal dimensions of the clusters.}
\label{dmf_p0.999999}
\end{center}
\end{figure}

In order to be more quantitative we calculate the
fractal dimension $d_{f}$. We measure the mass of the
invaded cluster for different generations $G$ for two
different probabilities $p^{*}=0.9$ and
$0.999999$. Numerical simulations were carried out for
$1000$ realizations on lattice sizes
$L=64,128,256$ and $512$. In Figs.~\ref{dmf_p0.9} and
\ref{dmf_p0.999999} we present log-log plots of the averaged
mass of the invaded cluster versus the lattice size
$L$. The linear fit to the data yields the fractal
dimension $d_{f}$ of the invaded cluster. In the case
$p^*=0.9$, the fractal dimension is $d_{f}=1.90\pm0.01$
for all generations. For $p^*=0.999999$ the fractal
dimension decreases when $G$ increases. This implies that
the fractal dimension of the invaded cluster has a behavior
similar to the previously studied case in which $p^*=1$.
%

%%%%%%%%%%%%%%%%%%%%%%%%%%%%%%%%%%%%%%%%%
\section{Conclusions}
%%%%%%%%%%%%%%%%%%%%%%%%%%%%%%%%%%%%%%%%%
We have presented a comprehensive model to study a multiple
invasion process. We have shown that the mass $M_{G}$ of the
invaded cluster decreases with the generation number $G$. In
addition, the fractal dimension of the invaded cluster
changes from $d_{1}=1.887\pm0.002$ to
$d_{s}=1.217\pm0.005$ corresponding to $G=1$
and $G=G_{s}$, respectively. This result confirms that the
multiple invasion process follows a continuous transition
from one universality class (NTIP) to another (optimal
path). We confirmed by extensive simulations that the
invaded mass follows a power law $M\sim G^{\beta}$ with an
exponent $\beta\approx0.6$. In addition the probability
distribution of avalanches $P(S,L,G)$ has been studied for
different system sizes as function of the parameter $G$. We
found that the mass distribution of avalanches follows a
power law where the exponent $\tau$ changes as function of
the generation number $G$. Based on this fact, we suggest
that the avalanche process belongs to a different {\it
universality class} for each $G$ since no crossover scaling
seems possible. Our results also indicate that this change
in universality class occurs in a continuous way. Concerning
the re-invasion of crystallizing, solidifying or healing
fluids we conclude that only in the case in which the
non-invaded part is not substantially damaged and the healed
parts typically do not get much stronger than they were
before the invasion, the multiple invasion process converges
well to a different universality class, namely that of the
optimal path
\cite{Schwarzer_99}. In the opposite case corresponding to
$p^{*}\neq 1$, the classical invasion percolation holds for
all generations.
%%%%%%%%%%%%%%%%%%%%%%%%%%%%%%%%%%%%%%%%%
\section{Acknowledgments}
%%%%%%%%%%%%%%%%%%%%%%%%%%%%%%%%%%%%%%%%%

We thank CNPq, CAPES, FINEP, FUNCAP, DFG Project $404$ and the Max Planck
prize for financial support.

\end{document}